# THE ENIGMA OF SATURN'S NORTH-POLAR HEXAGON

**Gerald E. Marsh**

**Argonne National Laboratory (Ret)**
gemarsh@uchicago.edu

**ABSTRACT**

It has been suggested that the north-polar hexagon found on Saturn is an unusual Rossby wave. If this is to be the case, one must not only explain how a Rossby wave can be hexagonal in shape, albeit with curved corners, but also why it is hexagonal rather than in the form of some other polygon. It is likely that a spectrum of Rossby waves with different amplitudes and wavelengths resulting from the velocity profile of the hexagonal jet is responsible for its shape.

**Introduction**

Saturn's north polar hexagon was discovered by Godfrey[1] in the Voyager spacecraft images in 1980-81 and has persisted for at least one Saturn year of about 29.5 Earth years. One of the images is shown in Fig. 1.

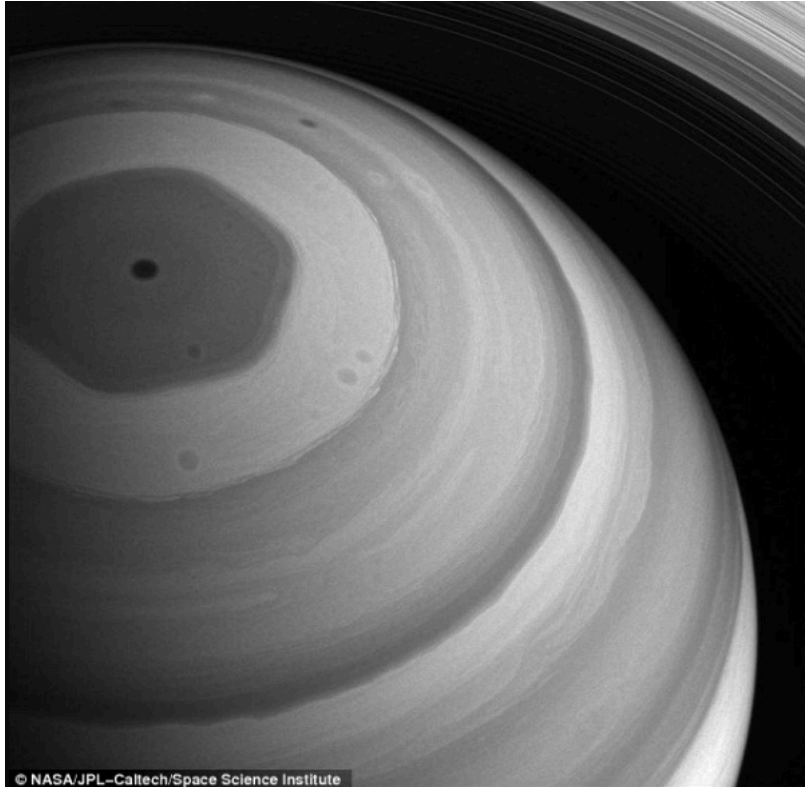

Figure 1. Saturn's north polar hexagon. Note the six well defined, albeit slightly curved, corners. The hexagon circles the north pole at a latitude of about 76° N. It is a cloud band associated with an eastward zonal jet having a peak velocity of about 100 m/s.

By tracking individual cloud patterns in the hexagon, the peak velocity of the eastward jet as has been measured to be about 100 m/s. Measurement of the velocity of the hexagon with respect to the planet—defined with respect to the reference frame† given by the Voyager-era radio period, which forms the basis for the currently defined System III rotation period—corresponds to an angular velocity of $\Omega_{III} = 1.64 \times 10^{-4}$ s$^{-1}$ in the westward direction, or about one revolution per Saturn day—this means that the hexagon is stationary with respect to fixed, non-rotating axes. Fletcher, et al.[2] found that the hexagon has a meridional temperature gradient with the equatorial side of the jet being colder than the polar side. This was measured at a depth into Saturn's atmosphere given by the pressure range 100-800 mbar. They suggested that this unexpected finding might be accounted for by an upwelling on the equatorial side of the polar jet and with a subsidence on the poleward side. Some dynamical explanation is needed since radiative heating near the north pole should be very limited. In any case, the hexagon itself appears to be insensitive to radiative heating since it persists over the large variations in radiative flux during the change in seasons.

---

† For the gas planets Saturn, Uranus, and Neptune, the rotation rate is taken to be that of their magnetic fields; this rotating coordinate system is known as System III. For Jupiter, there are two other systems known as System I—taken to be that of the average rotation rate of clouds in the equatorial region—and System II, for clouds outside the equatorial region.

A number of fluid-dynamical laboratory analogues that could be relevant to the study of Saturn's hexagonal jet have been performed. Perhaps the earliest are those done by Fultz in the 1950s.[3] An example is shown in Fig. 2.

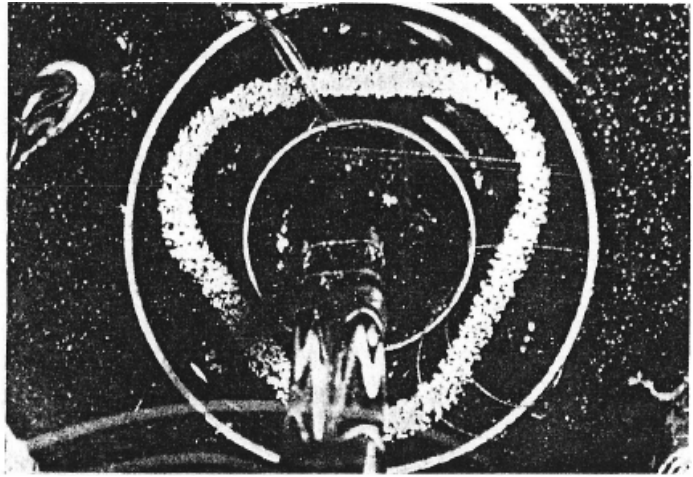

Figure 2. Water is rotated in the annulus between the two concentric cylinders. During the experiment the rotation rate was fixed and the temperature of the outer wall was slowly raised and the inner wall lowered keeping the mean temperature of the liquid constant. Aluminum powder on the surface collects in a continuous band showing the resulting analogue of a Rossby wave. [From reference 3]

As the temperature difference between the two concentric cylinders in Fig. 2 is raised above a critical point, the motion first transforms into a seven-wave pattern followed, as the temperature gradient increases, by lower number wave patterns (known today as the wave number—not to be confused with the zonal or hemispheric wave number, $k$, defined as $k = 2\pi r \cos\varphi/\lambda$, where $r$ is the radius of the planet, $\varphi$ is the latitude, and $\lambda$ is the wavelength). The three-wave pattern is shown in Fig. 2. This 3-lobed pattern is easy to reproduce analytically as seen in Fig. 3(a), and can be made to look more like Fig. (2), as can be seen in Fig. 3(b), by adding a shorter wavelength wave (this will be further discussed later).

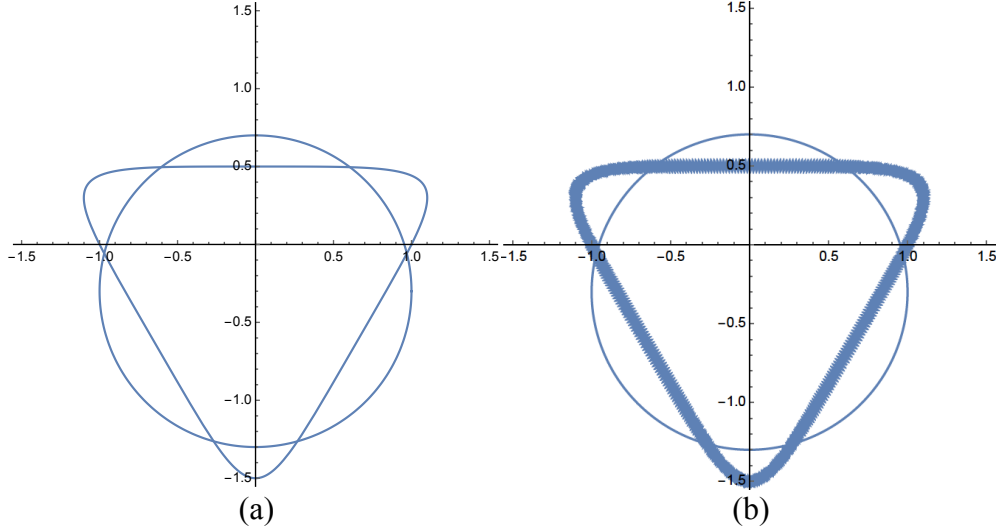

(a)          (b)

Figure 3. Analytic version of the 3-lobed Rossby wave pattern of Fig. 2. A circle with centroid or center of gravity at its origin. This offset can also be seen in Fig. 2. (a) The triangular figure is a parametric plot of $(1 + 0.5 \sin\theta)$ from 0 to $2\pi$. (b) The same with a shorter wavelength added to the parametric plot.

Using the parameters given by Fultz for Fig. 2, one can calculate $\lambda_s$, the stationary wavelength, using the equation given below [Eq. (2)]. With the dimensions given by Fultz, the wave number can then be calculated to be 2.98 ~ 3, which is what is seen in Fig. 2. This simple approach does not work for the 5-lobed pattern where other factors become involved. The difference between the 3 and 5-lobed flow patterns is the thermal Rossby number, which is given by Fultz as 0.12 and 0.058 respectively. The functional dependence of the dimensionless thermal Rossby number is given approximately by $R_{0_T} \sim \Delta T/\Omega^2$, $\Delta T$ being the temperature difference between the two cylindrical boundaries and $\Omega$ the rotation rate.

A simplified version of the wave-pattern transition curves, above the critical point found by Fultz, is shown in Fig. 4.

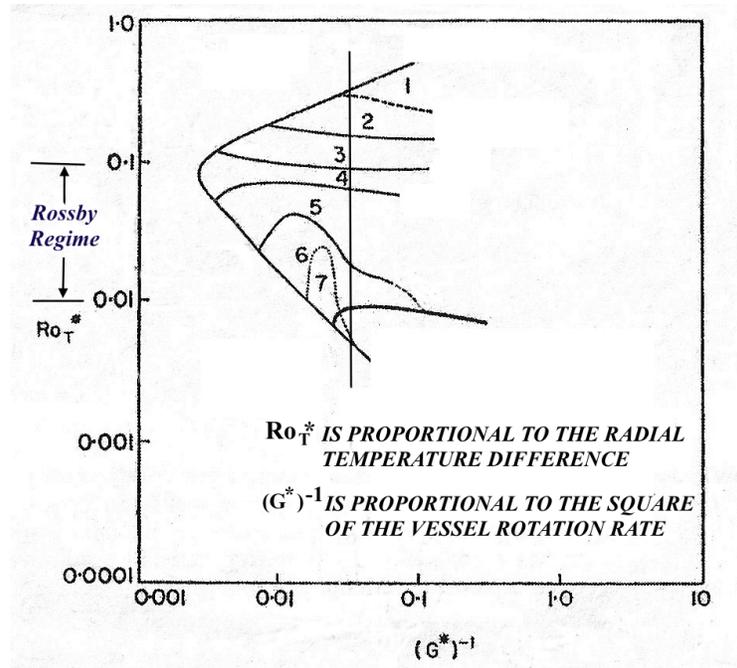

Figure 4. Simplified wave pattern transition curves. The variables were chosen to be dimensionless so as to make the results scale invariant. The vertical line corresponds to the series of experiments performed by Fultz. For small temperature differences, as is the case for Saturn, one would expect a six-wave pattern.

**The Hexagon**

Colwell[4] observed that a sinusoidal wave with a periodicity of six wavelengths could become a hexagon, albeit with curved corners, when wrapped around a circle. Adding a shorter wavelength has the effect of widening the hexagonal shape and making it closer to the observed hexagon. The amplitude of the shorter wavelength stationary wave determines the width of the pattern. Use of the parameters for Saturn's northern hexagon results in Fig. 5.

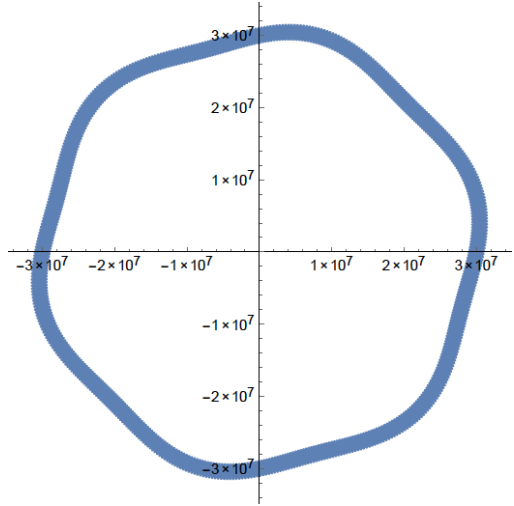

Figure 5. The hexagon obtained using the parameters for Saturn's northern hexagonal jet. The circumference of the latitude circle at $76^\circ$ north, where the center of the jet is located, is $9.4 \times 10^7 m$. The width of the jet (See Fig. 4) is set by the amplitude of the shorter wavelength wave; here it is $\sim 2\times 10^6$ $m$.

In Fig. 5, a single shorter wavelength of $1.9 \times 10^5 m$ was chosen to illustrate the principle. The velocity profile given in Eq. (6) tells us that a spectrum of shorter wavelengths with differing amplitudes would be expected to collectively form the actual jet

A more recent 2010 experiment was performed by Barbosa Aguiar, et al.[5] They found that in a zonally periodic domain regular stable polygons could form with an associated train of vortices. The flow remained vertically uniform so that the polygons and associated vortices extended through the whole depth of the fluid. Unlike the Fultz experiments, the bounding cylinders of the annulus were not heated or cooled. They also found that the zonal wave number depended on the wavelength that is most "energetically favored". Determining exactly what this means will be discussed in the summary.

Morales-Juberías, et al.[6] used the Explicit Planetary Isentropic-Coordinate General Circulation Model developed by Dowling, et al.[7] to simulate Saturn's hexagon. They used a Gaussian velocity distribution to represent the jet and found that the peak velocity and its curvature, which will be defined later, determined the dominant wave number.

**Rossby Waves**

It has been suggested that he north polar hexagon found on Saturn is an unusual Rossby wave. A Rossby wave is a meandering jet stream whose origin lies in a step like change in potential vorticity. They are transverse waves whose restoring force under a north or south displacement is proportional to the change in the Coriolis parameter.

Columns of air that decrease or increase their radius also respectively increase or decrease their vertical extent and relative vorticity. The potential vorticity is the ratio of the sum of the relative vorticity and Coriolis parameter to the height of the air column.

When warm air from the south meets the dense cold air from polar regions there is a pressure gradient between the cold polar air and the less dense warm air. This forms the polar front. This polar front has a step like decrease in the height of the tropopause along with a large lateral gradient in the potential vorticity across the step. In interaction with the Coriolis effect, this results in a high-speed jet stream flowing toward the east.

Perturbations of the jet stream can result in small-amplitude Rossby-like waves that can propagate along the jet and be trapped by it. That is, a tropopause level jet can be thought of as an enhanced potential vorticity gradient on an isentropic surface (a surface of constant entropy) that can serve as a waveguide for such perturbations. The concept of "waveguides" for planetary waves will be more fully discussed later in this paper.

**Rossby wave dispersion relation**

The simple model to be considered here is based on what in meteorology is called the Rossby wave dispersion relation and a Gaussian distribution for the jet velocity. For an atmosphere that can be considered to be an incompressible and horizontally homogenous fluid, the phase velocity of Rossby waves is given by

$$c = U - \frac{\beta \lambda^2}{4\pi^2},$$

(1)

where $U$ is the zonal velocity (here along the *x*-axis), $\lambda$ is the wavelength and the Rossby parameter, $\beta$, is given by $\beta = (2\omega \cos\varphi)/a$ (*a* is the radius of the planet). Rossby waves are meanders of the *upper* eastward circulation (known as high-altitude westerlies) and are always propagated eastward[†] relative to the medium and travel at speeds that depend on their wavelengths. Stationary waves can exist and their dispersion relation is given by setting $c = 0$ in Eq. (1), that is,

$$\lambda_s = 2\pi \left(\frac{U}{\beta}\right)^{1/2}.$$

(2)

This equation was used by Allison, et al. with $\beta$ put in terms of the gradient of the relative vorticity, $\varsigma_r = (\partial_x v - \partial_y u)$. The components of $U$ are ($u$, $v$, $w$). The barotropic relative vorticity equation can be written in the $\beta$-plane approximation, where the Coriolis parameter $f = 2\omega \sin\varphi$ varies linearly with latitude so that $f = f_0 + \beta y$, as

$$\frac{\partial \varsigma_r}{\partial t} + u\frac{\partial \varsigma_r}{\partial x} + v\frac{\partial \varsigma_r}{\partial y} + \beta v = 0,$$

(3)

where $\beta$ is here the meridional gradient of the Coriolis parameter; that is, $\beta = \partial_y f$. For the jet, where $\partial_x v = \partial_t \varsigma_r = 0$, Eq. (3) becomes

$$-\frac{\partial^2 U}{\partial y^2} + \beta = 0,$$

(4)

where $u$ can now be set equal to $U$ and $\partial^2 U / \partial y^2$ is now the relative vorticity gradient.[††] Substituting this result for $\beta$ into Eq. (2) gives

---

[†] Confusingly, an "eastwardly" wind blows westward and a "westerly" wind blows eastward.

[††] Some definitions: Potential vorticity is $q = (\varsigma_r + f)/h$, where $h$ is the thickness of a layer. For a constant thickness layer, the absolute vorticity is $\varsigma_a = \varsigma_r + f$, where $\varsigma_r$ is the relative vorticity and $f$ is the Coriolis parameter. The subscript *a* means relative to absolute space and the subscript *r* means relative to the Earth.

$$\lambda_s = 2\pi U^{1/2} \left(\frac{\partial^2 U}{\partial y^2}\right)^{-1/2}.$$

(5)

There are several values in the literature for the jet's relative vorticity gradient: Allison, et al.[8] average over the meridional *e*-folding interval to obtain $\approx 2.2 \times 10^{-11}$ $m^{-1}$ $s^{-1}$; that is, $\langle -\frac{\partial^2 U}{\partial y^2}\rangle_e \approx b/e = 2.2 \times 10^{-11} m^{-1} s^{-1}$, where *e* is the Naperian logarithm base 2.718… and *b* is peak latitudinal curvature defined in the next section. The plots in Del Genio, et al.[9] give a value of $-3 \times 10^{-11} m^{-1} s^{-1}$; and those in Aguiar, et al. a value of $-0.7 \times 10^{-11} m^{-1} s^{-1}$. For the peak velocity of the jet of 100 *m/s*, these correspond to a $\lambda_s$ of $1.34 \times 10^7 m, 1.13 \times 10^7 m$, and $2.38 \times 10^7 m$ respectively.

The circumference of the latitude circle at 76° north, the central position of the jet, is $9.4 \times 10^7 m$. If we divide this circumference by each of the above values for $\lambda_s$, the closest to the hexagon's number of six is that given by Allison, et al. of *n* = 7. If the values given by these authors are averaged the resulting $\lambda_s$ is $1.62 \times 10^7 m$, which results in *n* = 5.8.

**Gaussian profile for the Jet**

Allison, et al., as well as other authors, have used a Gaussian profile to represent the velocity of the hexagonal jet. The velocities are in the System III reference frame described above. The Gaussian profile is given by

$$U = U_0 e^{-by^2/2U_0}.$$

(6)

This profile, along with the Voyager data, is shown in Fig. 6. The peak velocity at the center of the jet is $U_0$ = 100 *m/s* and *y* is the meridional distance from the center of the jet.

The appearance of this type of velocity profile in Saturn's hexagonal jet is somewhat surprising for those familiar with fluid flow in circular pipes. A Gaussian velocity profile is usually thought of when one fluid is flowing into another. If buoyancy is not a factor, and only the fluid's momentum is important, the flow is called a jet. Such jets are turbulent. If the density of the fluid and jet are the same, and the cross section of the flow is circular, the envelope containing the turbulence caused by the jet is essentially conical with an angle measured from the axis of the jet to the conical envelope of 11.8 degrees. This angle is the same for all such jets. The velocity profile of this type of jet at different distances along the conical envelope is always a Gaussian, and these Gaussian velocity profiles are self-similar.[10]

There are also examples of jets that do not expand. On the Earth, for example, when cold, dense polar air meets the warmer and less dense air from the south a front is formed accompanied by a pressure gradient. This, under the influence of Coriolis forces, causes a strong flow aligned with the front. This flow forms a jet that does not expand linearly with distance along the jet; such upper atmosphere (troposphere) flows are called jet streams. They too have Gaussian velocity profiles.

An indication of whether the flow in a jet is turbulent or not can be determined by the velocity profile of the jet and the Rayleigh's necessary (but not sufficient) criterion for instability[11]: A necessary condition for instability is that the expression $\beta - \frac{\partial^2 U}{\partial y^2}$ change sign somewhere in the domain. A change in the sign of the vorticity is needed for instability.

The velocity profile of Fig. 6 below satisfies this necessary condition for instability, but the profile could be stabilized by the $\beta$-effect† if it is large enough since then $\beta - \frac{\partial^2 U}{\partial y^2}$ will not change sign. But, as discussed earlier, turbulence may not be a critical factor since even turbulent jets can have self-similar Gaussian profiles. Moreover, turbulence is not incompatible with the existence of large-scale, stable coherent structures, an example being the red spot of Jupiter.

---

† The "$\beta$-effect" is the systematic gradient of potential vorticity due to the spherical shape of the rotating planet and the meridional variation of the Coriolis parameter $\beta$ representing the variation at fixed latitude.

The peak latitudinal curvature of the jet in Fig. 6 is designated by b. The term "peak latitudinal curvature" comes from differential geometry where the tangent to a curve $x = x(s)$ is given by $t = dx(s)/ds$ and the "curvature" is $dt/ds = k(s)$; the curvature is generally defined as $|k(s)|$. Here, the curvature is $|-d^2U/dy^2|$. The "peak latitudinal curvature" is essentially the "peak" of the second derivative of the Gaussian curve modeling the jet. Explicitly, since the exponentials in Eq. (6) and the standard Gaussian must be the same, $b = U_0/\sigma^2$.

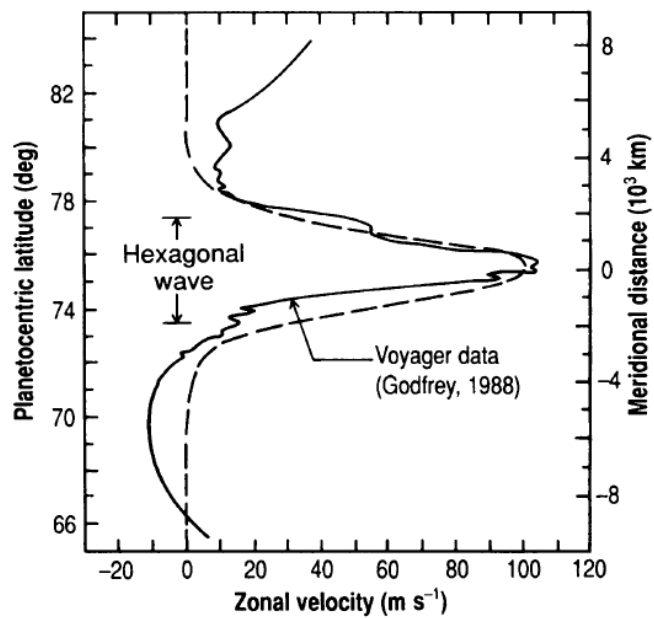

Figure 6. Zonal (latitudinal) velocity profile for the hexagonal jet. The dashed line is the Gaussian curve used to approximate the Voyager data. Note that the Gaussian profile satisfies the boundary condition $U = 0$ at the edges of the jet. Adapted from Allison, et al.

Because there is a maximum velocity, there is a maximum value for the wavelength $\lambda_s$. If one substitutes Eq. (6) into Eq. (2) and solves the resulting equation for $y$, the meridional distance from the center of the jet, its value becomes imaginary for $\lambda_s > 5.1 \times 10^7$ m. The longest wavelength allowed for a stationary wave, under the constraint that $\lambda_s$ divide the latitudinal circle at $76°$ north an integral number of times, is then $\lambda_s > 4.69 \times 10^7$ m, which corresponds to a wave number of 2. Other factors, not addressed here, select the higher $n = 6$ for the hexagon (See reference 7 for some discussion of this problem). The use of a velocity profile rather than a single velocity in Eq. (2) means that there is a spectrum of wavelengths $\lambda_s$ that compose the hexagon with, as found by Aguiar, et al., the zonal wave number depending on the wavelength that is most "energetically favored".

How the different wavelength waves in the jet combine is not really clear. To leading order, Rossby waves are linear and are generally explained in that context. Some work has been done, however, treating them as being weakly nonlinear.[12,13] It is thought that study of non-linear evolution equations for Rossby waves could help understand long-term coherent structures such as Saturn's hexagon, but this issue is beyond the scope of this paper.

One can define a two-dimensional stream function $\psi$ for Saturn's hexagon such that $u = -\partial_y \psi$ and $v = \partial_x \psi$, where the velocity $U = (u, v)$. Then the vorticity is the z-component of the relative vorticity $\zeta_r = (\nabla \times U)_z$. In terms of the stream function, $\zeta_r = \nabla^2 \psi$, where $\nabla^2$ is the two-dimensional Laplacian. Now the barotropic vorticity equation in a frame following the motion is Eq. (3) with $\partial_x \zeta_r = \partial_y \zeta_r = 0$ since $\zeta_r$ only has a z-component. In terms of the stream function Eq. (3) would then be

$$\frac{\partial}{\partial t}\nabla^2 \psi + \beta \frac{\partial \psi}{\partial x} = 0.$$

(7)

The dispersion relation given in Eq. (2) can be generalized to a Rossby wave traveling in an arbitrary direction by considering the wave to have the form $\psi = \exp[i(\vec{k} \cdot \vec{x} - \omega t)]$ where $\vec{k} = (k, l)$ and $\vec{x} = (x, y)$. Then, by requiring this form for $\psi$ to satisfy Eq. (7) one finds that

$$\frac{\omega}{k} = -\frac{\beta}{k^2 + l^2}.$$

(8)

The frequency can then be written as

$$\omega = -\frac{\beta \cos \alpha}{|\vec{k}|},$$

(9)

where the angle $\alpha$ is shown in Fig. 7, which plots the curves of constant $\omega$. In Fig. 7, the variables are $\alpha$ and $|\vec{k}|$.

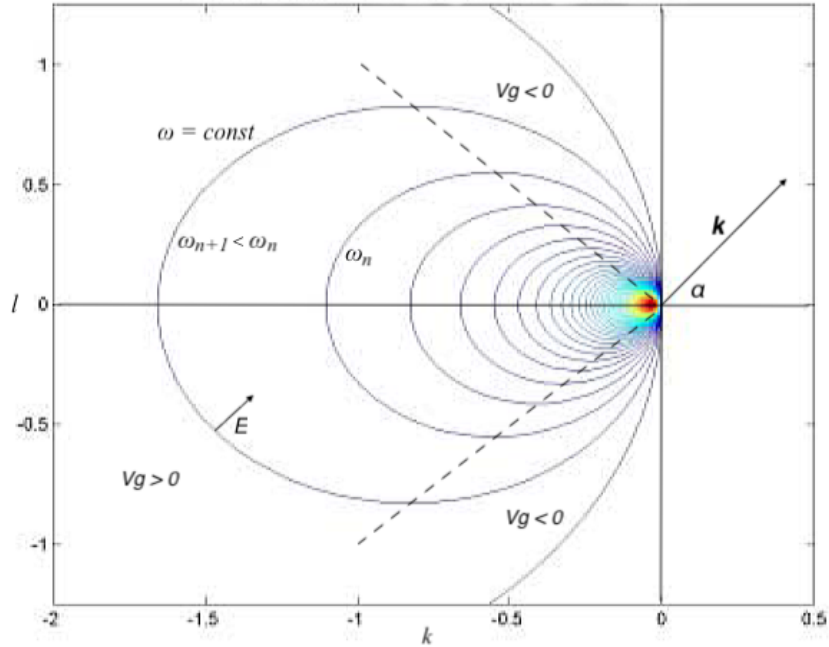

Figure 7. Curves of constant frequency $\omega$, with increasing frequency towards the origin. In contrast to non-dispersive waves, the wavelength is longer for higher frequency waves. The energy, $E$, carried by the waves is proportional to the group velocity $V_g^2$. $\mathbf{k}$ is the wave vector and $k$ is the component along the abscissa. The governing equation for the plot is $\omega = -\beta \cos \alpha /|\vec{k}|$ with $|\vec{k}| = \sqrt{k^2 + l^2}$. The orientation with respect to direction is that north is up and east is to the right. See the text for discussion of the group velocities and dashed lines. Note the aspect ratio, these curves are actually circles. [Adapted from Fig. 7 of the article in the *Encyclopedia of Atmospheric Sciences* (2002) by Peter B. Rhines, titled Rossby Waves.]

Saturn's northern hemisphere eastward jet has a concentrated gradient of relative vorticity at the core of the jet. The jet itself, because of the various velocities comprising the velocity profile of the jet, will have Rossby waves of different wavelengths associated with it. Collectively these Rossby waves will have a group velocity *Vg*, which *generally* defines the velocity of energy propagation. The group velocity points inward on the circles of constant $\omega$ shown in Fig. 7 toward the center of each circle.

The magnitude of the group velocity of Rossby waves is asymmetrical as can be seen by setting $x = k/l$ and recasting the frequency given in Eq. (8) as

$$\frac{\omega l}{\beta} = -\frac{x}{x^2 + 1},$$

(10)

and plotting the left hand side with respect to $x$ as shown in Fig. 8. Moreover, there are regions where the group velocity is *negative*.

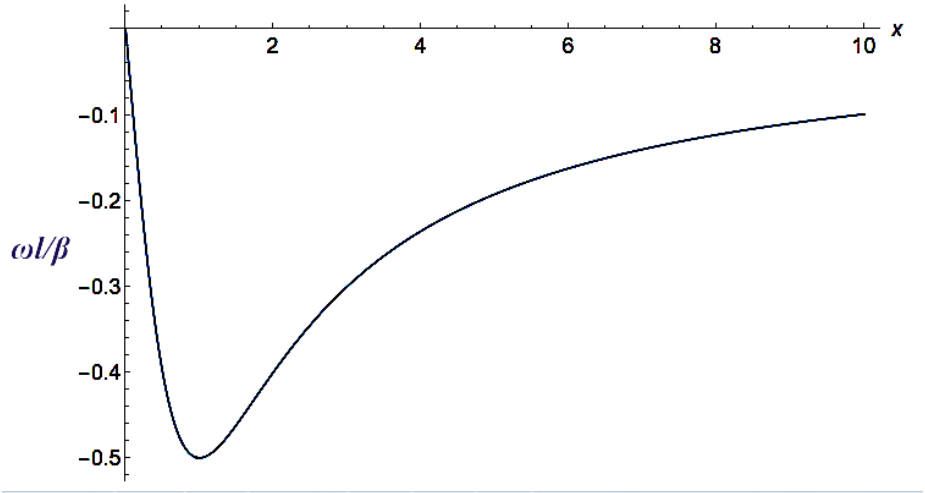

Figure 8. Plot of the frequency given by Eq. (10). The group velocity $\partial_k \omega$ corresponds to the tangent to the curve. Note that that the curve is asymmetrical and the tangent is negative for some range of $x$.

That regions exist where the group velocity must be negative can be seen by directly computing the group velocity from Eq. (9) to obtain

$$V_g = \frac{\partial \omega}{\partial k} = \frac{1}{k^2} \frac{\left(1 - \frac{l^2}{k^2}\right)}{\left(1 + \frac{l^2}{k^2}\right)^2}.$$

(11)

By setting $k = |\vec{k}| \cos \alpha$ and $l = |\vec{k}| \sin \alpha$ this equation can be written as

$$V_g = \frac{\partial \omega}{\partial k} = \frac{1}{|\vec{k}|^2 \cos^2 \alpha} \left( \frac{(1 - \tan^2 \alpha)}{(1 + \tan^2 \alpha)^2} \right).$$

(12)

The group velocity will then be negative whenever $\tan^2 \alpha > 1$; i.e., when $\pi/2 < \alpha < 3\pi/4$ and $5\pi/4 < \alpha < 6\pi/4$. These regions are indicated by the dashed lines in Fig. 7.

Negative group velocities in dispersive media often raise interpretational issues with regard to their apparent violations of causality: the peak of a transmitted pulse could exit the material before the peak of the incident pulse enters the material and in addition the pulse could propagate backward in the medium. However, it should be remembered that the signal velocity only coincides with the group velocity in regions of normal as opposed to anomalous dispersion. Experimentally, it has been shown that the peak of a pulse can indeed propagate in the backward direction, but the energy flow is always in the forward direction.[14] Causality is not violated. An extensive discussion of these issues has been given by Milonni.[15]

Saturn's hexagonal jet would be subject to perturbations that would be expected to affect the stationary Rossby waves presumed to compose it. The waves induced by the perturbation would be expected to have short wavelengths. The behavior of such waves has been discussed by Peter Rhines in his 2002 *Encyclopedia of Atmospheric Sciences* article. From the jet's velocity profile shown in Fig. 6, it can be seen that the flow has a shear on either side of the peak velocity. This shear will create a relative vorticity gradient. Rhines has argued that the "jet gains a concentrated gradient at the core" and that this will cause the short-wavelength waves due to the disturbance or perturbation to be trapped in the eastward jet, which then acts as a waveguide. The stability of the jet and its velocity profile would then be insensitive to small perturbations.

**Combining Rossby waves**

George Platzman,[16] in his 1968 Symons Memorial Lecture, introduced this waveguide analogy for planetary waves and Rossby waves in particular. Assuming a homogeneous, incompressible and inviscid atmosphere, and horizontal motion, he used the vorticity equation

$$\frac{d}{dt}\nabla^2\psi + 2\Omega\frac{\partial\psi}{\partial\lambda} = 0,$$

(13)

where $\psi$ is the stream function, $\nabla^2$ the surface spherical Laplacian, $\Omega$ the Earth's rotation speed and $\lambda$ the geographic longitude. The connection to the beta plane comes from $\nabla^2\psi$ being the vorticity and $2\Omega\ \partial\psi/\partial\lambda$ corresponding to $\beta v$. If one now assumes that the instantaneous distribution of $\psi$ over the sphere is a spherical harmonic, so that we are concerned with surface harmonics, then $\psi$ must satisfy

$$\nabla^2\psi + n(n+1)\psi = 0.$$

(14)

With a little algebra, Eq. (13), can be written as

$$\frac{\partial}{\partial t}\left(\nabla^2\psi - \frac{2\Omega}{\dot\lambda}\psi\right) = 0.$$

(15)

Following Longuett-Higgins[17] and comparing Eq. (15) to Eq. (14) tells us that $\psi$ will be a surface harmonic at an instant of time if

$$\dot\lambda = \frac{d\lambda}{dt} = -\frac{2\Omega}{n(n+1)}.$$

(16)

This means that the surface harmonic pattern will drift westward (in the direction opposite to the planetary rotation) with an angular velocity of $2\Omega/n(n+1)$ in geographic longitude. As mentioned earlier, the drift rate of Saturn's north-polar hexagon with respect to the planet is about one revolution per Saturn day. This means $n = 1$ for the hexagon and consequently, $\dot\lambda = -\Omega$. With respect to non-rotating axes, the hexagon is stationary.

**Jet streams as waveguides**
Let the surface harmonics be designated by $S_n^s$, where $s$ is the order and $n$ is the degree. These harmonics divide the sphere into blocks of alternate positive and negative values called tesserals. When $n = 0$, the sphere is divided into zonal harmonics where the alternate positive and negative values correspond to areas between two latitudes; and when $n = s$, the sphere is divided into sectors between two longitudes. Some examples are shown in Fig. 9.

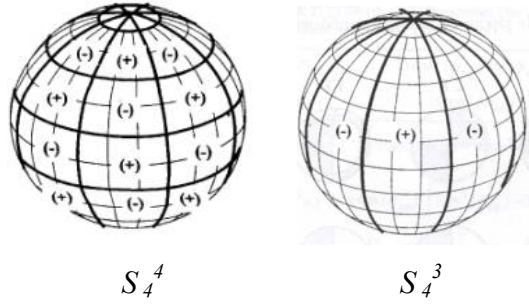

$S_4^4$   $S_4^3$

Figure 9. Examples of a tesseral harmonic and a sectoral harmonic.

Longuet-Higgins found that for waves constrained by two circles of latitude $\theta_1$ and $\theta_2$, the case that would be relevant for Saturn's hexagon, solutions to Eq. (13) must be of the form

$$\psi = [AP_n^s(\cos\theta) + BQ_n^s(\cos\theta)]e^{is(\phi-\dot\phi t)},$$

(17)

where $\dot\phi = -2\Omega/n(n+1)$. The constants $A$, $B$, and $n$ must be chosen so that $\psi$ vanishes when $\theta = \theta_1$ or $\theta_2$.

Platzman attributed a somewhat different form of Eq. (17) to Longuet-Higgins given by

$$\frac{1}{2}\left[P_n^s(\sin\theta) \pm \frac{2i}{\pi}Q_n^s(\sin\theta)\right]e^{is(\phi-\dot\phi t)}$$

(18)

and maintains that these functions add up to waveguide modes each of which has a sectorial, rather than tesseral, nodal configuration that can be considered the analogue of a homogeneous plane wave.

For Saturn's hexagon, $n = 1$ so that the only Legendre functions available are $P_1^1$ and $Q_1^1$. If $A$ and $B$ are constants in Eq. (17), it does not appear to be possible to satisfy the boundary condition that $\psi$ vanish when $\theta = \theta_1$ or $\theta_2$. Nor does it seem to be possible to obtain a Gaussian velocity profile for the flow between these boundaries.

The term "waveguide" is also used in a somewhat different context. Jet streams at the tropopause have a step-like decrease in the height of the tropopause and thus a strong lateral gradient of potential vorticity. Martius, et al.[18], refer to jet streams as waveguides and noted that such enhanced potential vorticity gradients on isentropic surfaces can serve as waveguides for perturbations.

**Summary**

In his early experiments, Fultz found that the wave number depended on the temperature gradient between the two cylinders constraining the fluid (see Fig. 2 and following discussion). The statement by Barbosa Aguiar, et al. that zonal wave numbers depend on the wavelength that is most "energetically favored" is difficult to define quantitatively because of various experimental factors. And despite the similarity parameters defined by Fultz[19], it is not certain how to confidently scale such laboratory experiments to Saturn, but such scaling does seem to work.

One might use the power spectra of the kinetic energy of the zonal wind profiles to help determine which zonal wave numbers are "energetically favored". This has been done for Jupiter by Choi and Showman.[20] They found that the smaller the wave number $k$ (that is, the longer the wavelength), the greater the spectral power. That this could be expected can be seen from Eq. (2), which implies that $U \propto \lambda_s^2$. Since the kinetic energy per unit mass is proportional to $U^2$, the energy is proportional to $\lambda_s^4$. This implies that the longer wavelengths would be energetically favored.

Based on their experiments, Barbosa Aguiar, et al. did find that Rossby numbers in the range of 0.03-0.1 favored the formation of hexagonal waves. This is consistent with Fultz's experiments where he found that a wave number of five corresponded to a Rossby number of 0.058, other conditions being the same as those for Fig.2.

Morales-Juberías, et al. found that the peak velocity of a Gaussian velocity distribution determined the dominant wave number. The model they used was initialized with a Gaussian velocity distribution for the jet whose peak velocity was 125 *m/s*. Their value for the curvature *b* in Eq. (6) was $10^{-10} m^{-1} s^{-1}$. After stabilization, the model showed a stable configuration having a wavenumber of six. When projected onto a polar map it has a hexagonal shape.

A discussion of the waveguide analogy for planetary and Rossby waves was given here and it was found that the boundary conditions for the solutions found by Longuet-Higgins and Platzman could not be satisfied for Saturn's hexagon where *n* = 1 in Eq. (14). On the other hand, if shorter wavelength waves are indeed trapped in a "waveguide", that could be the mechanism for combining different wavelength and amplitude Rossby waves to get the width of the jet as in Figs. 3 and 5.

Fultz's transition curves shown in Fig. 4 predicted that for a small temperature gradient the Rossby wave should take the form of a hexagon, and this is what we see on Saturn. Fultz wanted his results to be scale invariant, but could not possibly have imagined that sixty years later they would be scaled to the size of Saturn!

Thus, both questions asked in the Abstract would appear to have at least partially been answered. First by the straightforward geometrical considerations and elementary Rossby theory covered in this paper and second by the far more sophisticated model simulations of Morales-Juberías. The actual nature of the interaction of the waves comprising the jet is still open as is the issue of deriving the Gaussian velocity profile for the jet.